\documentclass[english,aps,prc,twocolumn,showpacs,superscriptaddress,nofootinbib,10pt]{revtex4-1} 
\usepackage[english]{babel}
\makeatletter
\let\l@English\l@english
\makeatother
\usepackage{graphicx}
\usepackage{capt-of}
\usepackage{amsmath}
\usepackage{amssymb}
\usepackage{relsize}
\usepackage[colorlinks,breaklinks]{hyperref}
\usepackage{xcolor}
\usepackage{multirow}
\newcommand{\bra}{\langle}
\newcommand{\ket}{\rangle}
\newcommand{\bs}[1]{\ensuremath{\boldsymbol{#1}}}
\newcommand{\nn}{\nonumber\\}

\newcommand{\be}{\begin{equation}}
\newcommand{\ee}{\end{equation}}
\newcommand{\bea}{\begin{eqnarray}}
\newcommand{\eea}{\end{eqnarray}}

\begin{document}

\title{Nuclear structure corrections to the Lamb shift in $\mu\,^3$He$^+$ and $\mu\,^3$H}

\author{N.~Nevo Dinur}
\email{nir.nevo@mail.huji.ac.il}
\affiliation{Racah Institute of Physics, The Hebrew University, Jerusalem 91904, Israel}

\author{C.~Ji}
\email{ji@ectstar.eu} 
\affiliation{TRIUMF, 4004 Wesbrook Mall, Vancouver, BC V6T 2A3, Canada}
\affiliation{ECT*, Villa Tambosi, 38123 Villazzano (Trento), Italy}
\affiliation{INFN-TIFPA, Trento Institute for Fundamental Physics and Applications, Trento, Italy}

\author{S.~Bacca}
\email{bacca@triumf.ca}
\affiliation{TRIUMF, 4004 Wesbrook Mall, Vancouver, BC V6T 2A3, Canada}
\affiliation{Department of Physics and Astronomy, University of Manitoba, Winnipeg, MB, Canada R3T 2N2}

\author{N.~Barnea}
\email{nir@phys.huji.ac.il}
\affiliation{Racah Institute of Physics, The Hebrew University, Jerusalem 91904, Israel}

\date{\today}

%%%%%%%%%%%%%%%%%%%%%%%%%%%%%%%%%%%%%%%%%%%%%%%%%%%
%%%%%%%%%%%%%%%%%%%%%%%%%%%%%%%%%%%%%%%%%%%%%%%%%%%

\begin{abstract}
Measuring the \mbox{2S-2P} Lamb shift in a hydrogen-like muonic atom 
allows one to extract its nuclear charge radius with a high precision 
that is limited by the uncertainty in the nuclear structure corrections.
The charge radius of the proton thus extracted 
was found to be $7\sigma$ away from the CODATA value, 
in what has become the yet unsolved ``proton radius puzzle''.
Further experiments currently aim at the isotopes of hydrogen and helium: 
the precise extraction of their radii may provide a hint at the solution of the puzzle.
We present the first {\it ab~initio} calculation of nuclear structure corrections, 
including the nuclear polarization correction, to the \mbox{2S-2P} transition in
$\mu\,^3$He$^+$ and $\mu\,^3$H, and assess solid theoretical error bars.
Our predictions reduce the uncertainty in the nuclear structure corrections
to the level of a few percents and will be instrumental to the on-going $\mu\,^3$He$^+$ experiment. 
We also support the mirror $\mu\,^3$H system as a candidate for further 
probing of the nucleon polarizabilities
and shedding more light on the puzzle.
\end{abstract}


\maketitle

%%%%%%%%%%%%%%%%%%%%%%%%%%%%%%%%%%%%%%%%%%%%%%%
% Introduction
%%%%%%%%%%%%%%%%%%%%%%%%%%%%%%%%%%%%%%%%%%%%%%%
\section{Introduction}\label{sec:Intro}
The root-mean-square (RMS) charge radius of the proton 
$r_p \equiv \sqrt{\bra r_p^2 \ket}$ 
was recently determined 
with unprecedented precision 
from laser spectroscopy measurements of \mbox{2S-2P} transitions  
in muonic hydrogen $\mu$H, where the electron is replaced by a muon~\cite{Pohl:2010_Nat,Antognini:2013_Sci}.  
The extracted $r_p$  differs by 7$\sigma$ from the CODATA value~\cite{CODATA2010}, 
 which is  based  in turn  on many measurements involving electron-proton interactions.
This discrepancy  between the `muonic' and `electronic' proton radii ($r_p(\mu^-)$ and $r_p(e^-)$, respectively)
is known as the ``proton radius puzzle,'' and has attracted much attention (see, {\it e.g.}, Ref.~\cite{Pohl:2013_ARNPS} for an extensive review and Ref.~\cite{Pohl:2014_Hyper} for a brief summary of current results and ongoing experimental effort). 
In an attempt to solve the puzzle, 
extractions of $r_p(e^-)$ from the ample electron-proton ($ep$) 
scattering data have been reanalyzed by, {\it e.g.}, Refs.~\cite{Hill:2010_PRD,Sick:2014_PRC-1,Kraus:2014_PRC,Lorenz:2015_PRD}, 
while several planned experiments aim to remeasure $ep$ scattering in new kinematic regions 
relevant for the puzzle~\cite{JLAB_ep_lowQ2,Mainz_ISR}. 
$r_p$ extracted from electronic hydrogen is also being reexamined, 
both theoretically~\cite{Karshenboim:2015_PRA} 
and experimentaly~\cite{Vutha:2012_BAPS,Beyer:2013_AdP,Peters:2013_AdP}, 
as well as the Rydberg constant~\cite{Tan:2011_PS,Peters:2013_AdP}, 
which is relevant for several radius extraction methods. 
A few of the theoretical attempts to account for the discrepancy between 
$r_p(e^-)$ and $r_p(\mu^-)$
include new interactions that violate lepton universality~\cite{Batell:2011_PRL,Tucker-Smith:2011_PRD,CarlsonRislow:2014_PRD} 
and novel proton structures~\cite{Hill:2011_PRL,Birse:2012_EPJA,Miller:2013_PLB,Jentschura:2013_PRA,Hagelstein:2015_PRA}. 
Yet the puzzle has not been solved. 
Answers may be provided (see, {\it e.g.} Refs.~\cite{Miller:2011_PRA,Tomalak:2014_PRD}) 
by a planned experiment at PSI~\cite{MUSE-Gilman:2013_AIPCP} to scatter electrons, muons, and their antiparticles off the proton using the same experimental setup.

Alternatively, it will be insightful to study 
whether the puzzle also exists in other light nuclei, 
and whether it depends on the atomic mass $A$, charge number $Z$, or the number of neutrons $N$.
In particular, the CREMA collaboration plans to extract high-precision charge radii 
from Lamb shift measurements that were recently performed 
in several hydrogen-like muonic systems~\cite{Antognini:2011_Can,Pohl:2014_Hyper}, 
namely, $\mu$D, $\mu\,^3$He$^+$, and $\mu\,^4$He$^+$.
These measurements may unveil a dependence of the discrepancy on the isospin of the measured nucleus and, 
in particular, probe whether the neutron exhibits a similar effect as the puzzling proton. 
To obtain some control over these issues, it is advisable that nuclei with different $N/Z$ ratios will be mapped out. 
It is the purpose of this Letter to perform an {\it ab~initio} calculation of nuclear structure 
corrections (including nuclear polarization), with solid error estimates, for the  $\mu\,^3$He$^+$ system and for its nuclear mirror, $\mu\,^3$H.

The Lamb shift~\cite{Lamb:1947_PR} is the \mbox{2S-2P} energy difference consisting of
\begin{equation} \label{eq:E2s2p}
\Delta E \equiv \delta_{\rm QED} 
+ \delta_{\rm FS}\!\left(R_c\right) 
+ \delta_{\rm TPE} 
~ , 
\end{equation}
where,  in decreasing order of magnitude, the three terms include: 
quantum electro-dynamics (QED) contributions from vacuum polarization, lepton self-energy, and relativistic recoil in $\delta_{\rm QED}$; 
finite-nucleus-size contributions in $\delta_{\rm FS}\!\left(R_c\right)$, 
where $R_c\equiv\sqrt{\bra R^2_c\ket}$ is the nuclear RMS charge radius;
and contributions from two-photon exchange (TPE) between the lepton and the nucleus in $\delta_{\rm TPE}$.
The last term can be divided into the elastic Zemach term and the inelastic polarization term, 
{\it i.e.}, $\delta_{\rm TPE} = \delta_{\rm Zem}+\delta_{\rm pol}$. 
Additionally, each of these terms is separated into contributions from nuclear ($\delta^{A}$) and nucleonic ($\delta^{N}$) degrees of freedom, 
$\delta_{\rm TPE} = \delta_{\rm Zem}^{A}+\delta_{\rm Zem}^{N}+ \delta_{\rm pol}^{A}+\delta_{\rm pol}^{N}$.

In light muonic atoms, 
$\delta_{\rm QED} \approx 10^2$--$10^3$~meV and is estimated from theory with a precision better than $10^{-3}$~meV~\cite{Borie:2012_AoP,Borie-arXiv:1103_2014,Krauth-arXiv:1506,Krutov:2015_JETP}.
At leading order $\delta_{\rm FS}\!\left(R_c\right)=\frac{m_r^3 (Z\alpha)^4}{12} R_c^2$, 
with $m_r$ the reduced mass of the muon-nucleus system, 
while higher-order contributions are at the sub-percentage level~\cite{Borie:2012_AoP}. 
The limiting factor for the attainable precision of $R_c$ extracted from Eq.~\eqref{eq:E2s2p} is by far the uncertainty in $\delta_{\rm TPE}$.  
This was confirmed in two recent papers that reviewed the theory in $\mu$D~\cite{Krauth-arXiv:1506}, 
and in $\mu\,^4$He and $\mu\,^3$He~\cite{Krutov:2015_JETP}. 
Ref.~\cite{Krauth-arXiv:1506} 
covers all the theoretical contributions to the Lamb shift in $\mu$D, 
including a summary of recent efforts by several groups~\cite{Pachucki:2011_PRL, Hernandez:2014_PLB, Carlson:2014_PRA, Pachucki:2015_PRA} to accurately obtain $\delta_{\rm TPE}$ in $\mu$D and reliably estimate its uncertainty, which comes out an order of magnitude larger than the uncertainties in the other terms. 
Ref.~\cite{Krutov:2015_JETP} details all the contributions for the two helium isotopes. 
Many terms are recalculated, not including the polarization correction $\delta_{\rm pol}$. 
For $\mu\,^4$He$^+$, {\it ab~initio} nuclear calculations
were recently applied in Ref.~\cite{Ji:2013_PRL}, 
improving on decades-old estimates of $\delta_{\rm pol}$.
For three-body nuclei, the only calculations of $\delta_{\rm pol}$ are outdated; 
based on old and simplistic nuclear models, their results are either inaccurate~\cite{Joachain:1961_NP} or imprecise~\cite{Rinker:1976_PRA}, reinforcing the need for modern, accurate, {\it ab~initio} calculations for the three-body nuclei.

%%%%%%%%%%%%%%%%%%%%%%%%%%%%%%%%%%%%%%%%%%%%%%%
% Calculation details
%%%%%%%%%%%%%%%%%%%%%%%%%%%%%%%%%%%%%%%%%%%%%%%
\section{Nuclear Structure Contributions}\label{sec:Calc_details}
The nuclear Zemach term $\delta^A_{\rm Zem}$ enters Eq.~\eqref{eq:E2s2p} as the elastic nuclear-structure contribution to $\delta^A_{\rm TPE}$\footnote{
	$\delta^A_{\rm Zem}$ was derived by Friar~\cite{Friar:1979_AoP} as the first-order $Z\alpha$ correction to 
	$\delta_{\rm FS}(R_c)$ and is called the `Friar' term in Ref.~\cite{Krauth-arXiv:1506}.}.
This term is of order $(Z\alpha)^5$ and is defined as
\begin{equation}
\label{eq:del_Zem_def}
\delta^A_{\rm Zem} = -\frac{m_r^4(Z \alpha)^5}{24} \langle r^3\rangle_{(2)}~, 
\end{equation}
where 
$\langle r^3\rangle_{(2)}$ is the 3rd nuclear Zemach moment\footnote{
	\label{foot:Zemach}
	We refer only to charge-charge Zemach moments; for more details see, {\it e.g.}, Ref.~\cite{Distler:2011_PLB}.}. 
Friar \& Payne showed~\cite{FriarPayne:1997_PRA} 
that the first-order corrections in $\delta^A_{\rm pol}$ contain a part that cancels $\delta^A_{\rm Zem}$ exactly. 
Calculation of this part can thus be avoided, as was done in Ref.~\cite{Pachucki:2011_PRL}, 
providing only the sum $\delta^A_{\rm TPE} = \delta^A_{\rm pol} + \delta^A_{\rm Zem}$. 
However, following Refs.~\cite{Ji:2013_PRL,Ji:2014_FBS,Hernandez:2014_PLB}, 
we calculate explicitly all the parts of $\delta^A_{\rm pol}$, 
including the Zemach term, as detailed below.
This is done in order to:
(a)~allow comparison with other values in the literature, 
and 
(b)~provide theoretical support for the alternative way of extracting $R_c$ from Eq.~\eqref{eq:E2s2p} 
where the Zemach term is phenomenologically parameterized as~\cite{Borie:2012_AoP} 
\begin{equation}
\label{eq:C_factor}
\delta_{\rm Zem}^{A}=\mathcal{C} \times R_c^3~. % C = -1.35(4) meV fm$^{-3}$ for 3He in Ref.~\cite{Borie:2012_AoP}
\end{equation}

%%%%%%%%%%%%%%%%
% polarization
%%%%%%%%%%%%%%%%
 As in Refs.~\cite{Ji:2013_PRL,Hernandez:2014_PLB}, 
 the energy correction due to nuclear polarization is obtained as a sum of contributions
\begin{align}\label{eq:delta_A_pol_terms}
\delta^{A}_{pol} &= 
\left[\delta^{(0)}_{D1} + \delta^{(0)}_{L} + \delta^{(0)}_{T} + \delta^{(0)}_{C}+ \delta^{(0)}_{M} \right]
+ \left[\delta^{(1)}_{R3} + \delta^{(1)}_{Z3} \right]
\nn
&\quad 
+ \left[\delta^{(2)}_{R^2} + \delta^{(2)}_{Q} + \delta^{(2)}_{D1D3} \right]
+ \left[\delta^{(1)}_{NS} + \delta^{(2)}_{NS} \right]. 
\end{align}
Detailed formulas pertaining to most of the terms in Eq.~\eqref{eq:delta_A_pol_terms} 
are found in~\cite{Ji:2013_PRL} and are not repeated here. 
The largest contribution comes from the leading term, $\delta^{(0)}_{D1}$, related to the electric dipole. 
To this we add relativistic longitudinal and transverse corrections $\delta^{(0)}_{L}$ and $\delta^{(0)}_{T}$, respectively, as well as Coulomb distortion corrections $\delta^{(0)}_{C}$. 
Here we follow Ref.~\cite{Hernandez:2014_PLB} 
and include in $\delta^{(0)}_{C}$ only the logarithmically enhanced term from the next order in $Z\alpha$.
We generalize the treatment in Ref.~\cite{Hernandez:2014_PLB} of the magnetic term $\delta^{(0)}_M$ by using the impulse approximation operator that includes the orbital angular momentum \cite{Ring80}.
First-order corrections 
$\delta^{(1)}_{R3}$ and $\delta^{(1)}_{Z3}$
are related to a proton-proton correlation term and to the 3rd nuclear Zemach moment, respectively.
Finally, at the next order we have the 
monopole $\delta^{(2)}_{R^2}$, quadrupole $\delta^{(2)}_{Q}$, and interference $\delta^{(2)}_{D1D3}$ terms.
All the above terms are calculated using point nucleons. 
Finite-nucleon-size (NS) corrections appear in Eq.~\eqref{eq:delta_A_pol_terms} 
as $\delta^{(1)}_{NS}=\delta^{(1)}_{R1}+\delta^{(1)}_{Z1}$ and $\delta^{(2)}_{NS}$, which we elaborate on below.

%%%%%%%%%%%%%%%%%%%%%%%%%%%%%%%%%%%%%%%%%%%%%%%
% NS Corrections
%%%%%%%%%%%%%%%%%%%%%%%%%%%%%%%%%%%%%%%%%%%%%%%
\section{Nucleon-Size Corrections}\label{sec:NS Corrections}
%-----------------------------------------------
The TPE in the point-nucleon limit is expressed as the interaction of photons 
with the structureless charged protons, while the neutrons are ignored. 
In this limit, the point-proton density operator is 
\begin{equation}
\label{eq:rhoR_p}
\hat{\rho}_p(\bs{R}) \equiv 
\frac{1}{Z}\sum_{a=1}^A \delta(\bs{R}-\bs{R}_a)\frac{1+\tau_a^3}{2}~,
\end{equation}
where $\tau_a^3$ is the isospin projection operator.
When the finite nucleon sizes are considered, 
$\hat{\rho}_p(\bs{R})$ must be convoluted with the proton's internal charge distribution,  
and a similar convolution is applied to the point-neutron density operator 
\begin{equation}
\label{eq:rhoR_n}
\hat{\rho}_n(\bs{R})  \equiv 
\frac{1}{N}\sum_{a=1}^A \delta(\bs{R}-\bs{R}_a) \frac{1-\tau_a^3}{2}~.
\end{equation}
Following Refs.~\cite{Ji:2013_PRL,Ji:2014_FBS}, 
we apply a low-momentum expansion for the nucleon form factors, 
parameterized here by their mean square charge radii, 
$r^2_{n/p} \equiv \bra r^2_{n/p} \ket$. 
We adopt $r_n^2 = {-0.1161(22)}$~fm$^2$~\cite{PDG_Beringer:2012_PRD}. 
For the proton, we may use either 
\mbox{$r_p(e^-) = 0.8775(51)$}~fm~\cite{CODATA2010} or 
\mbox{$r_p(\mu^-) = 0.84087(39)$}~fm~\cite{Antognini:2013_Sci}. 
In fact, until the ``proton radius puzzle'' is resolved 
(or when $R_c$ and other properties of the nuclei under consideration are measured using muons), 
we should use 
$r_p(e^-)$ 
for comparison with the literature, 
which is based on data obtained with electrons, 
and 
$r_p(\mu^-)$ 
for predictions in muonic systems.

The leading NS correction $\delta^{(1)}_{NS}$ is the sum of nucleon-nucleon correlations in 
$\delta^{(1)}_{R1}$ and Zemach-like terms in $\delta^{(1)}_{Z1}$. 
%%%%%%%%%%%%%%%
%% Correlations
%%%%%%%%%%%%%%%
The former is expressed as
\begin{align}\label{eq:delta_1_NS}
\delta^{(1)}_{R1}
=& {-\frac{m_r^4(Z \alpha )^5}{6}} 
\iint d\bs{R}d\bs{R}' |\bs{R}-\bs{R}'|
\Big[ 
r^2_p \bra 0| \hat{\rho}_p^\dagger(\bs{R})\hat{\rho}_p(\bs{R}')| 0\ket
\nn
+&
\frac{N}{Z}
r^2_n \bra 0| \hat{\rho}_n^\dagger(\bs{R})\hat{\rho}_p(\bs{R}')| 0\ket
\Big]~, 
\end{align}
which includes proton-proton ($pp$) and neutron-proton ($np$) correlations. 
It is an NS correction to the point-nucleon contribution $\delta^{(1)}_{R3}$ of Eq.~\eqref{eq:delta_A_pol_terms} 
(the latter is denoted $\delta^{(1)}_{R3pp}$ in Ref.~\cite{Ji:2013_PRL}).
%%%%%%%%%%%%
%% Zemach
%%%%%%%%%%%%
For the calculation of Zemach-like terms using point-nucleons we define
\begin{equation}
\langle r^k_{ij}\rangle_{(2)} \equiv \iint d \bs{R} d \bs{R}' \left|\bs{R}-\bs{R}'\right|^k 
\bra 0|\hat{\rho}_i^\dagger(\bs{R})|0\ket\bra 0| \hat{\rho}_j(\bs{R}')|0\ket~,
\end{equation}
with $i$, $j$ denoting either $p$ or $n$. 
The 3rd nuclear Zemach moment is thus calculated as 
\begin{equation}
\label{eq:3rd_Zem}
\langle r^3\rangle_{(2)} %= 
\approx
\langle r^3_{pp}\rangle_{(2)} 
+4 \left[ 
r^2_p \langle r^1_{pp}\rangle_{(2)}
+
\frac{N}{Z} r^2_n \langle r^1_{np}\rangle_{(2)}
\right]~,
\end{equation}
where the first term is the point-nucleon limit and the second is the (approximated) NS correction. 
Accordingly, the point-nucleon Zemach term $\delta^{(1)}_{Z3}$ and its NS correction $\delta^{(1)}_{Z1}$ 
are obtained by inserting Eq.~\eqref{eq:3rd_Zem} into Eq.~\eqref{eq:del_Zem_def} and flipping the sign, 
{\it i.e.}, $\delta^A_{\rm Zem}  \approx -( \delta^{(1)}_{Z3}+\delta^{(1)}_{Z1} )$.

%%%%%%%%%%%%%%%
%% 2nd order NS
%%%%%%%%%%%%%%%
The sub-leading NS correction $\delta^{(2)}_{NS}$ is evaluated through 
a sum rule of the dipole response~\footnote{The sign before $r^2_n$ in Eq.~\eqref{eq:delta_2_NS} is corrected from Refs.~\cite{Ji:2013_PRL,Hernandez:2014_PLB} and agrees with Ref.~\cite{Pachucki:2015_PRA}.}
\begin{equation}
\label{eq:delta_2_NS}
\delta^{(2)}_{NS} = {-\frac{8\pi}{27}} m_r^5 (Z \alpha )^5 
\left[ r^2_p - \frac{N}{Z} r^2_n\right]
\int^\infty_{\omega_{\rm th}}
d\omega \sqrt{\frac{\omega}{2m_r}} S_{D_1}(\omega)~.
\end{equation}

%%%%%%%%%%%%%%%%%%%%%%%%%%%%%%%%%%%%%%%%%%%%%%%%%%%%%%%%%
Lastly, the nucleonic TPE correction $\delta^N_{\rm TPE}$ also enters Eq.~\eqref{eq:E2s2p}. 
We defer the treatment of this hadronic contribution to a dedicated section below.

%%%%%%%%%%%%%%%%%%%%%%%%%%%%%%%%%%%%%%%%%%%%%%%
% Methods
%%%%%%%%%%%%%%%%%%%%%%%%%%%%%%%%%%%%%%%%%%%%%%%
\section{Methods}\label{sec:Methods}
%-------------------------
Most of the above contributions can be written as sum rules of several nuclear responses with various energy-dependent weight functions~\cite{Ji:2013_PRL,Hernandez:2014_PLB}. 
They were evaluated using the newly developed Lanczos sum rule method~\cite{NevoDinur:2014_PRC}.
Ground-state observables of $^3$He and $^3$H, 
as well as Lanczos coefficients, 
were obtained using the effective interaction hyperspherical harmonics (EIHH) method~\cite{Barnea:2000_PRC,Barnea:2001_NPA}.
As only ingredients we employed in the nuclear Hamiltonian either one of the following state-of-the-art nuclear potentials: 
(i) the phenomenological AV18/UIX potential with two-nucleon~\cite{AV18} plus three-nucleon~\cite{UIX} forces; and 
(ii) the chiral effective field theory $\chi$EFT 
potential with two-nucleon~\cite{Entem:2003_PRC} plus three-nucleon~\cite{Navratil:2007_FBS} forces.

%{\it Error Estimates} ---
It is of utmost importance to have realistic uncertainty estimates for our nuclear TPE predictions. 
These terms are the least well known in Eq.~\eqref{eq:E2s2p}, 
and their uncertainties determine the attainable precision of $R_c$ extracted from Lamb shift measurements.
We considered many sources of uncertainty, 
 namely: 
 numerical; 
 nuclear model; 
 isospin symmetry breaking; 
 higher-order nucleon-size corrections; 
 missing relativistic and Coulomb-distortion corrections; 
 higher multipoles, % eta-expansion
 terms of higher-order in $Z\alpha$; 
 and the effect of meson-exchange currents on the magnetic contribution. 
Their individual and cumulative effect on $\delta^A_{\rm pol}$, $\delta^A_{\rm Zem}$, and $\delta^A_{\rm TPE}$ have been estimated and applied to the results given below. 
More details about these uncertainty estimates are given in the Supplementary Materials~\cite{NevoDinur:2015_MuA3_supp}.

%%%%%%%%%%%%%%%%%%%%%%%%%%%%%%%%%%%%%%%%%%%%%%%
% Results
%%%%%%%%%%%%%%%%%%%%%%%%%%%%%%%%%%%%%%%%%%%%%%%
\subsection{Results}\label{sec:Results}
%-------------------------
We first compare a few observables we have calculated for the $^3$He and $^3$H 
nuclei with corresponding theoretical and experimental values available in the literature.
In Table~\ref{table_checks} we present the ground-state 
binding energy BE, 
charge radius $R_c$, 
 and
magnetic moment $\hat{\mu}_{gs}$, 
as well as 
 the electric dipole polarizability $\alpha_E$. 
In general, good agreement is found with other calculations.

Our results do not include isospin-symmetry breaking (ISB), 
except for the Coulomb interaction between protons in $^3$He. 
Calculations by other groups shown in Table~\ref{table_checks} usually do not include ISB effects;
notable exceptions are Ref.~\cite{Kievsky:2008_JPG}, 
which includes the $T=3/2$ isospin channel in the ground-state wave function, 
and Ref.~\cite{Nogga:2003_PRC} that provides results either including or excluding it. 
One observes that including ISB alters BE by a few keV. 
In addition, the $^3$He BE, not used in the calibration of the Hamiltonians, 
is overestimated at a sub-percentage level, and this is slightly {\it worsened} when ISB is included.
As discussed in Ref.~\cite{Stetcu:2009_PRC}, changes in BE shift the threshold of sum rules, 
affecting mostly sum rules with inverse energy dependence, such as $\alpha_E$ discussed below. 
For the other observables in Table~\ref{table_checks}, the estimated uncertainty stemming from ISB is $\lesssim$1\%.

\begin{table}[htb]
\caption{ Various $^3$He and $^3$H observables (see text for details)
	calculated with the AV18/UIX and $\chi$EFT
	potentials, compared to corresponding calculations in the literature 
	and to experimental values. 
	 Our ground-state wave functions do not include the $T=3/2$ channel.
	 Our numerical uncertainties are not shown since they are smaller than one in the last decimal place.
	References labels correspond to:
$^{a/b}$~Ref.~\cite{Nogga:2003_PRC} without/with inclusion of the $T=3/2$ %isospin 
channel, respectively;
$^c$~Ref.~\cite{Kievsky:2008_JPG} 
(which includes the $T=3/2$ channel); 
$^d$~Ref.~\cite{Purcell:2010_NPA};
$^e$~Ref.~\cite{Marcucci:2008_PRC};
$^f$~Ref.~\cite{Pastore:2013_PRC};
$^g$~Ref.~\cite{Pachucki:2007_PRA},
$^h$~Ref.\cite{Leidemann:2003_FBSS};
$^i$~Ref.~\cite{Stetcu:2009_PRC}; 
$^j$~Ref.~\cite{Rinker:1976_PRA};
$^k$~Ref.~\cite{Goeckner:1991_PRC};
$^l$~Ref.~\cite{Efros:1997_PLB};
$^m$~Ref.~\cite{Angeli:2013_ADNDT}. 
}
\label{table_checks}
\begin{center}
\footnotesize
\renewcommand{\tabcolsep}{2.mm}
\begin{tabular}{@{}l l l l  l@{}} %c}
\hline\hline
{\bf $\bs{^3}$He}
			&BE~[MeV]    &$R_{c}(e^{-}) $~[fm] &$\hat{\mu}_{gs}$~[$\mu_N$] &$\alpha_E$~[fm$^{3}$]\\
 \hline
AV18/UIX    
			&7.740       &$\ $1.968      &${-1.73}$                 &0.149\\ 
Lit.
			&7.740(1)$^a$&$\ \quad$-     &${-1.764}^e$     		    &0.153(15)$^g$\\
			&7.748(1)$^b$&$\ \quad$-     &${-1.749}^f$  			&0.145$^h$\\
 \hline
$\chi$EFT
			&7.735       &$\ $1.988      &${-1.76}$				    &0.153\\ 
Lit.
			&7.750$^c$   &$\ \quad$-     &$\quad$-     			    &0.149(5)$^i$\\
\hline
 Exp.
			&7.71804$^d$ &$\ $1.966(3)$^m$ &${-2.127}^d$            &0.130(13)$^j$\\
			&            &             &                            &0.250(40)$^k$\\
\hline \hline
% Triton
{\bf $\bs{^3}$H}%&    &       &              & \\
			&BE~[MeV]    &$R_{c}(e^{-})$~[fm] &$\hat{\mu}_{gs}$~[$\mu_N$] &$\alpha_E$~[fm$^{3}$]\\
\hline
AV18/UIX
			&8.473       &$\ $1.755      &2.59 	      			    &0.137\\ 
Lit.
			&8.472(1)$^a$&$\ \quad$-     &2.575$^e$        		    &0.139(4)$^l$\\
			&8.478(1)$^b$&$\ \quad$-     &2.569$^f$	     		    &\\
\hline
$\chi$EFT 
			&8.478       &$\ $1.777      &2.63            		    &0.139\\ 
Lit.
			&8.474$^c$   &$\ \quad$-     &$\quad$-     			    &0.139(2)$^i$\\
\hline
Exp.
			&8.48180$^d$ &$\ $1.759(36)$^m$& 2.979$^d$        		   &\\
\hline \hline
\end{tabular}
\end{center}
\end{table}
%%%%%%%%%%%%%%%%%%%%%%%%%%%%%%%%%%%%%%%%%%%%%%%%%%%%%%%%%%%%%%%%%%%%%%%%%%%%%

%%%%%%%%%%%%%%%%%%%%%%%%%%
Charge radii $R_c$ shown in Table~\ref{table_checks} 
are obtained from the calculated point-proton-distribution 
RMS radius $R_p$ as~\cite{Friar:1997_PRA,SO-Radius_Ong:2010_PRC}
\begin{equation}
\label{eq:Rc_radii_relation}
R_c^2 =  R_p^2 + r_p^2 + \frac{N}{Z} r_n^2 + \frac{3}{4 m^2_p}
~, %~. 
\end{equation}
 where we omit contributions from the spin-orbit radius (negligible for s-shell nuclei) and meson-exchange currents.
The last term in Eq.~\eqref{eq:Rc_radii_relation} is the Darwin-Foldy term, 
where $m_p$ is the proton mass, taken from Refs.~\cite{CODATA2010,PDG_Beringer:2012_PRD}. 
In Table~\ref{table_checks}, 
we show only $R_c$ values obtained using $r_p(e^-)$ and experimental values obtained only with electrons. 
As a direct result of Eq.~\eqref{eq:Rc_radii_relation}, using $r_p(\mu^-)$ would decrease $R_c$ by 0.016 (0.018) fm for $^3$He ($^3$H). 
We note that the uncertainty, currently governed by nuclear-model dependence, 
is slightly larger than the effect of varying $r_p$.
It should also be noted that our $R_p$ values 
agree with the hyperspherical harmonics calculations of the Pisa group~\cite{Kievsky:2008_JPG} 
for both nuclear potentials, suggesting a small ISB effect, 
while the Monte-Carlo calculations of Ref.~\cite{Pastore:2013_PRC} show less agreement and hint at a larger ISB effect. 
Considering that radii were not included in the calibration of the nuclear Hamiltonians, 
it would be interesting to further investigate their sensitivity to the theoretical apparatus. 
In particular, work is in progress to include meson-exchange currents~\cite{Pastore:prep}. 
%%%%%%%%%
Currently, 
for $^3$He the AV18/UIX charge radius 
is in better agreement with the experimental value, 
while for $^3$H 
the experimental error bar is larger than the nuclear-model dependence,
and calls for a more precise measurement.

%%%%%%%%%%%%%%%%%%%%%%%%%%
Concerning the magnetic moments, 
our results are comparable with the other impulse approximation calculations presented in Table~\ref{table_checks}, 
which deviate from experiment due to the absence of meson-exchange currents. 
However, we do not include meson-exchange currents in $\delta^A_{\rm TPE}$, 
since the contribution of the magnetic term $\delta^{(0)}_M$ is small enough to make these corrections negligible.

%%%%%%%%%%%%%%%%%%%%%%%%%%
The electric dipole polarizability $\alpha_E$ is an inverse-energy-weighted sum rule of  
 the dipole response and is therefore closely related to $\delta^A_{\rm pol}$.
Our results are in agreement with previous calculations, especially the recent Ref.~\cite{Stetcu:2009_PRC}. %, which uses a $\chi$EFT potential. 
As in~\cite{Ji:2013_PRL}, $\alpha_E$ is found to be nuclear-model dependent.
We provide first results for the unmeasured $\alpha_E$ of $^3$H with the AV18/UIX potential, which lies within the uncertainty estimates of~\cite{Stetcu:2009_PRC}.

% Zemach moments
%=================
We now turn to the 
Zemach terms, 
first listing available values in the literature. 
%
% Borie:
In Refs.~\cite{Borie:2012_AoP,Borie-arXiv:1103_2014} Borie 
calculated $\delta^A_{\rm Zem}$, following Friar~\cite{Friar:1979_AoP}, 
using a Gaussian distribution that fits the nuclear-charge-radius obtained from electron experiments.
The result\footnote{
	\label{foot:Borie}
	Ref.~\cite{Borie-arXiv:1103_2014} is the arXiv version of Ref.~\cite{Borie:2012_AoP}, which has been updated since publication; in particular, $\delta^A_{\rm Zem}\left(^3{\rm He}\right)$ was increased by $\sim$20\% with respect to the published version. 
}$^,$\footnote{
	\label{foot:sign}The result is given using our sign convention.}
was 
\mbox{$\delta^A_{\rm Zem}\left(^3{\rm He}\right)={-10.258(305)}$~meV}. 
%
% Krutov:
Recently, Krutov et al.~\cite{Krutov:2015_JETP} repeated this calculation and obtained 
\mbox{$\delta^A_{\rm Zem}\left(^3{\rm He}\right)={-10.28(10)}$~meV}. 
% 
% Sick:
Alternatively, 
inserting the 3rd nuclear Zemach moment 
recently extracted from $e-^3$He scattering data~\cite{Sick:2014_PRC-2} 
into Eq.~\eqref{eq:del_Zem_def} gives 
\mbox{$\delta^A_{\rm Zem}\left(^3{\rm He}\right) ={-10.87(27)}$~meV}. 
As explained above, 
all of these results should be compared 
with our calculation that uses $r_p(e^-)$ as input and yields
$\delta^A_{\rm Zem}\left(^3{\rm He}\right)\left[ r_p(e^-)\right] = {-10.71(19)(16)}$~meV, 
where the first uncertainty results from nuclear-model dependence and the second includes all other sources. 
Our result is in agreement with these references 
	(based on comments made in Refs.~\cite{Sick:2014_PRC-2,Krutov:2015_JETP}, 
	we assume that the error-bars in Ref.~\cite{Krutov:2015_JETP} are not exhaustive). 
However, for the muonic systems considered here 
 we use $r_p(\mu^-)$, which gives 
\begin{equation}
\label{eq:del_Zem_result_3He}
\delta^A_{\rm Zem}\left(^3{\rm He}\right)\left[ r_p(\mu^-)\right]={-10.49(19)(16)}~{\rm meV}. 
\end{equation}
We note that with the given error-bars this result is also in agreement with Refs.~\cite{Sick:2014_PRC-2,Borie-arXiv:1103_2014,Krutov:2015_JETP}.

% C-factor
The use of Eq.~\eqref{eq:C_factor} is adopted from Refs.~\cite{Borie:2012_AoP,Borie-arXiv:1103_2014}, 
where\footnote{See footnote~\ref{foot:sign}.} $\mathcal{C}\left(^3{\rm He}\right)=-1.35(4)$ meV fm$^{-3}$. 
The results of Ref.~\cite{Sick:2014_PRC-2} can also be used to extract 
$\mathcal{C}\left(^3{\rm He}\right)=\delta_{\rm Zem}^{A}/R_c^3=-1.42(4)$ meV fm$^{-3}$ 
from the $e-^3$He scattering data. 
Our calculations of $\delta_{\rm Zem}^{A}$ and $R_c$ with either value of $r_p$ give 
$\mathcal{C}\left(^3{\rm He}\right)\left[ r_p(e^-)\right] = -1.383(05)(20)$ meV fm$^{-3}$ and 
$\mathcal{C}\left(^3{\rm He}\right)\left[ r_p(\mu^-)\right] = -1.388(05)(21)$ meV fm$^{-3}$, 
which both agree with 
Refs.~\cite{Borie:2012_AoP,Borie-arXiv:1103_2014,Sick:2014_PRC-2}. 
Evidently, the nuclear-model dependence is diminished for this value, 
since it is proportional to the geometrical ratio 
$\langle r^3\rangle_{(2)} / R_c^3$. 
Similarly to $R_c$ discussed above, 
the difference between $\delta_{\rm Zem}^{A}$ results 
obtained with the two nuclear potentials stems from the different point-proton distributions, 
and this largely cancels out in $\mathcal{C}$, 
reducing its total relative uncertainty compared to $\delta_{\rm Zem}^{A}$.

% 3H
Repeating the above procedures we obtain predictions for $\mu\,^3$H 
\begin{eqnarray}
\label{eq:del_Zem_result_3H}
% \delta^A_{\rm Zem}\left(^3{\rm H}\right)\left[ r_p(e^-)\right] &=& {-0.233(5)(3)}~{\rm meV} \nn 
\delta^A_{\rm Zem}\left(^3{\rm H}\right)\left[ r_p(\mu^-)\right] &=& {-0.227(5)(3)}~{\rm meV}, 
\end{eqnarray}
and
\begin{eqnarray}
\label{eq:C_result_3H}
% \mathcal{C}\left(^3{\rm H}\right)\left[ r_p(e^-)\right] &=& {-0.0423(2)(6)}~{\rm meV~fm}^{-3} \nn 
\mathcal{C}\left(^3{\rm H}\right)\left[ r_p(\mu^-)\right] &=& {-0.0425(2)(6)}~{\rm meV~fm}^{-3}. 
\end{eqnarray}
For future comparisons, 
using $r_p(e^-)$ shifts $\delta^A_{\rm Zem}\left(^3{\rm H}\right)$ by ${-6}\ \mu$eV 
and $\mathcal{C}\left(^3{\rm H}\right)$ by ${+0.2}\ \mu{\rm eV~fm}^{-3}$.

%%%%%%%%%%%%%%%%%%%%%%%%%%%%%%%%%%%
% Nuclear polarization corrections
%-------------------------------
%\subsection{Nuclear polarization}\label{subsec:Nuclear_pol_results}

%%%%%%%%%%%%%%%%%%%%%%%%%%
\setkeys{Gin}{width=\textwidth}
\begin{figure}[hbt]
	\centering{
		{\includegraphics[angle=0,clip=true,trim=140 60 60 140 %amounts to remove from the left, bottom, right and top
			,width=\linewidth]
			{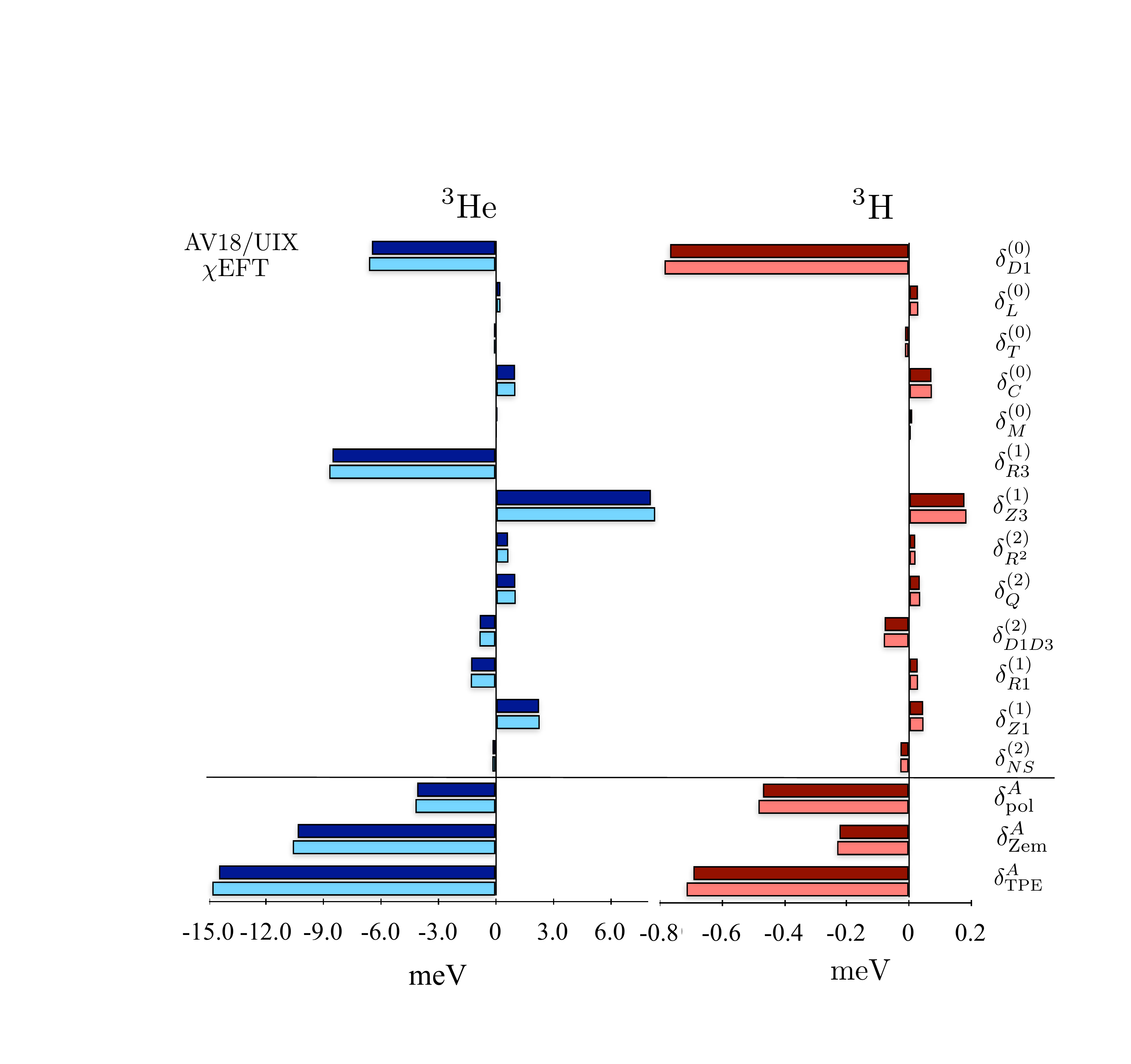}}}
	\caption{
		Graphic representation of the various contributions 
		to the nuclear structure and polarization corrections to the \mbox{2S-2P} Lamb shift 
		in the muonic hydrogen-like systems of $^3$He and $^3$H, 
		calculated with the AV18/UIX and $\chi$EFT nuclear potentials. 
		Notice the different scales used for the two systems.
	} 
	\label{fig_delta_pol_A} 
\end{figure}
%%%%%%%%%%%%%%%%%%%%

Next, the nuclear polarization correction to the Lamb shift --- $\delta^A_{\rm pol}$ --- is obtained by summing up the terms in Eq.~\eqref{eq:delta_A_pol_terms}. 
Their values for $\mu\,^3$He$^+$ and $\mu\,^3$H, calculated with the two nuclear potentials, 
are shown\footnote{
	The numerical values are detailed in the Supplementary Materials~\cite{NevoDinur:2015_MuA3_supp}.} 
in Fig.~\ref{fig_delta_pol_A}. 
Here, the NS corrections are obtained using only $r_p(\mu^-)$. 
Taking the mean value of the two nuclear potentials 
we obtain 
\begin{eqnarray}
\label{eq:delta_A_pol_results}
%\delta^A_{\rm pol} \left( \mu\,^3{\rm He}^+\right) &=& {-4.157(253)}\ {\rm meV}\nn
%\delta^A_{\rm pol} \left( \mu\,^3{\rm H }  \right) &=& {-0.476(15)}\ {\rm meV}~.
\delta^A_{\rm pol} \left( \mu\,^3{\rm He}^+\right) &=& {-4.16(06)(16)}\ {\rm meV}\nn %tot err=(24.9)
\delta^A_{\rm pol} \left( \mu\,^3{\rm H }  \right) &=& {-0.476(10)(13)}\ {\rm meV}~, %tot err=(13)
\end{eqnarray}
where we retain the use of first and second brackets for uncertainties from nuclear-model dependence and from all other sources, respectively.
 Our result for $\mu\,^3$He$^+$ agrees with Rinker's ${-4.9}\pm1.0$ meV obtained fourty years ago~\cite{Rinker:1976_PRA}. 
The $\mu\,^3$H case was rarely studied.
We note, however, that a comparison with the simplistic calculation of Ref.~\cite{Joachain:1961_NP} 
reveals a similar ratio of $\sim$9 between $\delta^A_{\rm pol}$ of $\mu\,^3$He$^+$ and of $\mu\,^3$H, both in Ref.~\cite{Joachain:1961_NP} and in our work.

Adding Eqs.~\eqref{eq:del_Zem_result_3He} and \eqref{eq:del_Zem_result_3H} 
to Eq.~\eqref{eq:delta_A_pol_results} we obtain the total nuclear-structure 
TPE corrections that enter Eq.~\eqref{eq:E2s2p} 
\begin{eqnarray}
\label{eq:delta_A_TPE_results}
%\delta^A_{\rm TPE} \left( \mu\,^3{\rm He}^+\right) &=& {-14.64(38)}\ {\rm meV}\nn
%\delta^A_{\rm TPE} \left( \mu\,^3{\rm H }  \right) &=& {-0.703(17)}\ {\rm meV}~.
\delta^A_{\rm TPE} \left( \mu\,^3{\rm He}^+\right) &=& {-14.64(25)(27)}\ {\rm meV}\nn %tot err=(36.5)
\delta^A_{\rm TPE} \left( \mu\,^3{\rm H }  \right) &=& {-0.703(16)(11)}\ {\rm meV}~. %tot err=(19)
\end{eqnarray}

%%%%%%%%%%%%%%%%%%%%%%%%%%%%%%%%%%%%%%%%%%%%%%%
\section{Hadronic TPE}\label{sec:Nucleon_pol}
%-------------------------
The last ingredient in $\delta_{\rm TPE}$ 
is the contribution from two-photon exchange with the internal degrees of freedom of the 
nucleons that make up the nucleus, {\it i.e.}, $\delta^N_{\rm TPE} = \delta^N_{\rm Zem} + \delta^N_{\rm pol}$.  
Since it is dictated by the hadronic scale, about 10 times higher than the nuclear interaction, 
this contribution can be approximated as the sum of TPE effects with each of the individual nucleons. 
The various terms that contribute to $\delta^N_{\rm TPE}$ are estimated based on 
previous studies of $\mu$H, as recently done for $\mu$D in Ref.~\cite{Krauth-arXiv:1506}. 
Specifically, as suggested by Birse and McGovern~\cite{BirseMcGovern:pc}, 
we adopt values of $\delta_{\rm Zem}$ and $\delta_{\rm pol}$ in $\mu$H that are combinations of results from Refs.~\cite{Carlson:2011_PRA,Birse:2012_EPJA}, as detailed below.

%%%%%%%%%%%%%%%%%%%%%%
% Nucleon Zemach
%%%%%%%%%%%%%%%%%%%%%%
As Friar showed in Ref.~\cite{Friar:2013_PRC}, 
the intrinsic Zemach term of each proton contributes to $\delta_{\rm TPE}$ of the nucleus as an additional NS correction, not accounted for in the NS corrections detailed above\footnote{
	In our notations this term appears as an NS correction to $\delta^{(1)}_{R3}$.}. 
We denote this term $\delta^N_{\rm Zem}$ and find its contribution to be proportional to the 
 analogous term in $\mu$H by 
\begin{equation}
\label{eq:new_Proton_Zemach_term}
\delta^N_{\rm Zem} (\mu {\rm A})
= \left( \frac{Z m_r(\mu {\rm A})}{m_r(\mu {\rm H})}\right)^4 \times \delta_{\rm Zem}(\mu {\rm H})~. 
\end{equation} 
We take $\delta_{\rm Zem}(\mu{\rm H})=0.0247(13)$ meV~\footnote{
	We use the same value as in~\cite{Krauth-arXiv:1506}. 
	Here, $\delta_{\rm Zem}(\mu{\rm H})$ stands for the elastic + non-pole parts of $\delta_{\rm TPE}(\mu{\rm H})$, 
	and not for the non-relativistic limit that is related to the proton's 3rd Zemach moment 
	(see Refs.~\cite{Antognini:2013_Supp,Antognini:2013_AoP}).} 
and obtain 
\begin{eqnarray}
\label{eq:delta_N_Zem_results}
\delta^N_{\rm Zem} \left( \mu\,^3{\rm He}^+\right) &=& {-0.487(26)}\ {\rm meV}\nn
\delta^N_{\rm Zem} \left( \mu\,^3{\rm H }  \right) &=& {-0.0305(16)}\ {\rm meV}~.
\end{eqnarray}

%%%%%%%%%%%%%%%%%%%%%%
% Nucleon polarization
%%%%%%%%%%%%%%%%%%%%%%
In Ref.~\cite{Carlson:2014_PRA}, $\delta^N_{\rm pol}$ of $\mu$D was extracted from electron scattering data. 
Here, we resort to estimating $\delta^N_{\rm pol}$ by relating it to the proton polarization correction in $\mu$H via~\cite{Pachucki:2011_PRL,Miller-arXiv:1501,Pachucki:2015_PRA} 
\begin{equation}
\label{eq:pol-N}
\delta^N_{\rm pol}(\mu {\rm A}) = (N+Z) \left[Z m_r(\mu {\rm A})/m_r(\mu {\rm H})\right]^3 \delta_{\rm pol}(\mu {\rm H})~, %.
\end{equation}
assuming that the neutron polarization contribution is the same as that of the proton.
Here we use $\delta_{\rm pol}(\mu{\rm H})=9.3(1.1)$ $\mu$eV\footnote{
	$\delta_{\rm pol}(\mu {\rm H})=\delta^p_{\rm inelastic} + \delta^p_{\rm subtraction}$.
	For the former we follow Ref.~\cite{Carlson:2014_PRA} and adopt 13.5 $\mu$eV, 
	which is an average of three values from Ref.~\cite{Carlson:2011_PRA}, 
	and for the latter we use  ${-4.2(1.0)}$ $\mu$eV from Ref~\cite{Birse:2012_EPJA}.}. 
Based on current knowledge of the nucleon polarizabilities~\cite{Myers:2014_PRL}, 
we assign an additional 20\% uncertainty to the neutron polarization contribution.
Another possible error in $\delta^N_{\rm pol}$ arises from neglecting 
medium effects and nucleon-nucleon interferences in Eq.~\eqref{eq:pol-N}.
These effects can be estimated by comparing the calculated $\delta^N_{\rm pol}(\mu{\rm D})$ 
with the result evaluated in Ref.~\cite{Carlson:2014_PRA} from scattering data.
This yields a $\sim$29\% correction. 
Until this correction is calculated rigorously in other light muonic atoms, 
we estimate it to be of a similar size, multiplied by $A/2$, 
making it the dominant source of uncertainty in our $\delta^N_{\rm TPE}$. 
Eventually, we obtain 
\begin{eqnarray}
\label{eq:delta_N_pol_results}
\delta^N_{\rm pol} \left( \mu\,^3{\rm He}^+\right) &=& {-0.275(123)}\ {\rm meV}\nn
\delta^N_{\rm pol} \left( \mu\,^3{\rm H }  \right) &=& {-0.034(16)}\ {\rm meV}~.
\end{eqnarray}

%%%%%%%%%%%%%%%%%%%%%%
% Nucleon TPE
%%%%%%%%%%%%%%%%%%%%%%
Summing up the results in Eqs.~\eqref{eq:delta_N_Zem_results} and \eqref{eq:delta_N_pol_results} we obtain 
the total contribution from the nucleon degrees of freedom 
\begin{eqnarray}
\label{eq:delta_N_TPE_results}
\delta^N_{\rm TPE} \left( \mu\,^3{\rm He}^+\right) &=& {-0.762(125)}\ {\rm meV}\nn
\delta^N_{\rm TPE} \left( \mu\,^3{\rm H }  \right) &=& {-0.065(16)}\ {\rm meV}~.
\end{eqnarray}
In $\mu\,^3 {\rm He}^+$, $\delta^N_{\rm TPE}$ is $\sim$5\% of $\delta^A_{\rm TPE}$, 
{\it i.e.}, about twice the overall uncertainty in $\delta^A_{\rm TPE}$. 
For $\mu\,^3$H we obtained that $\delta^N_{\rm TPE}$ is $\sim$9\% of $\delta^A_{\rm TPE}$, 
which is more than three times the uncertainty in the latter. 
Therefore, our precision in predicting $\delta^A_{\rm TPE}$ can be important 
not only for the determination of $R_c$ from muonic Lamb shift measurements, 
but also for probing $\delta^N_{\rm TPE}$, 
if these measurements reveal discrepancies with electronic experiments 
that may indicate exotic contributions to $\delta^N_{\rm TPE}$.
A study of the Lamb shift in $\mu\,^3$H will be especially sensitive to the nucleon polarizabilities, 
since their relative contribution is much larger in this case.

%%%%%%%%%%%%%%%%%%%%%%%%%%%%%%%%%%%%%%%%%%%%%%%
% Conclusion
%%%%%%%%%%%%%%%%%%%%%%%%%%%%%%%%%%%%%%%%%%%%%%%
\section{Summary}\label{sec:Summary}
%-------------------------
We have performed the first {\it ab~initio} calculation of 
$\delta_{\rm Zem}^{A}$ and $\delta^A_{\rm pol}$
for both $\mu\,^3$He$^+$ and $\mu\,^3$H, 
using state-of-the-art nuclear potentials. 
Many possible sources of uncertainty have been considered, 
yet the resulting uncertainties of a few percents are much lower than in previous estimates of 
$\delta^A_{\rm pol}$ and $\delta^A_{\rm TPE}$, 
which relied on imprecise data and simplistic models. 
In addition, our $\delta_{\rm Zem}^{A}$ calculations agree with previous estimates 
and with recent analysis of $e^-\,^3$He scattering, 
and provide predictions towards $^3$H measurements. 
They were also adapted for muonic systems by incorporating $r_p(\mu^-)$ --- the proton radius measured with muons.

Ultimately, 
our results will allow two alternative ways of extracting a much more precise $R_c$ 
from a recent measurement~\cite{Antognini:2011_Can,Pohl:2014_Hyper,Kottmann:pc} of the Lamb shift in $\mu\,^3$He$^+$, 
and from an analogous measurement we encourage to conduct in $\mu\,^3$H.
The precision of the charge radii of $^3$He and $^3$H could be thus improved by factors of $\sim$5 and $\sim$50, respectively, which could have interesting implications for nuclear physics.

Finally, we estimate the hadronic contribution 
$\delta^N_{\rm TPE}$ 
in these systems, and find it to be larger than our uncertainty estimates in $\delta^A_{\rm TPE}$. 
Therefore, this combined theoretical and experimental effort may not only shed some light on the 
``proton radius puzzle,'' but could also probe the elusive nucleon polarizabilities tightly connected to it.

%%%%%%%%%%%%%%%%%%%%%%%%%%%%%%%%%%%%%%%%%%%%%%%
\begin{acknowledgments}
NND would like to thank Bezalel Bazak for suggesting the case of muonic triton, 
to express a special thanks to the Mainz Institute for Theoretical Physics (MITP) for its hospitality and support, 
and acknowledge constructive discussions with 
Aldo Antognini, Mike Birse, Michael Distler, Mikhail Gorchtein, 
Savely Karshenboim, Franz Kottmann, Randolf Pohl, 
and Ingo Sick.
This work was supported in parts by 
the Natural Sciences and Engineering Research Council (NSERC), 
the National Research Council of Canada, 
the Israel Science Foundation (Grant number 954/09),
and the Pazy foundation.
\end{acknowledgments}

%%%%%%%%%%%%%%%%%%%%%%%%%%%%%%%%%%%%%%%%%%%%%%%%%%%%%%%
% Bibliography
%%%%%%%%%%%%%%%%%%%%%%%%%%%%%%%%%%%%%%%%%%%%%%%%%%%%%%%

\bibliographystyle{apsrev4-1}
%\bibliography{muA3_bib}

%\bibliography{muA3_bib.bib}
% 1. import bibitems from below into: "muA3_bib.bib"
% 2. add missing bibitems to "muA3_bib.bib"
% 3. create "*.bbl" file and copy its contents instead of what's below

%%%%%%%%%%%%%%%%%%%%%%%%%%%%%%%%%%%%%%%%%%%%%%%%%%%%%%%%%%%%%%%%%%%%%%%%
\appendix
%%%%%%%%%% Supplementary Materials %%%%%%%%%%
\pagebreak
\widetext
\begin{center}
\textbf{\large Supplementary Materials}
\end{center}
%%%%%%%%%% Commands for supplementary materials %%%%%%%%%%
%%%%%%%%%% Prefix a "S" to all equations, figures, tables and reset the counter %%%%%%%%%%
\setcounter{equation}{0}
\setcounter{figure}{0}
\setcounter{table}{0}
\setcounter{page}{1}
\makeatletter
\renewcommand{\theequation}{S\arabic{equation}}
\renewcommand{\thefigure}{S\arabic{figure}}
\renewcommand{\thetable}{S\arabic{table}}
\makeatother
%%%%%%%%%% Prefix a "S" to all equations, figures, tables and reset the counter %%%%%%%%%%

\section{error estimation}
We consider many sources of uncertainty. 
Below we explain the origin and derivation of each uncertainty estimate. 
\begin{description}
	\item [Numerical] 
	First we estimate the numerical accuracy of the calculations. 
	In the EIHH method~\cite{Barnea:2000_PRC,Barnea:2001_NPA}, the calculations are usually repeated with the model space truncated at increasing values of the maximal hyperangular momentum $K_{\rm max}$, until the differences between consecutive $K_{\rm max}$ results become negligible. 
	Accordingly, 
	the numerical uncertainty can be estimated, 
	as in~\cite{Ji:2013_PRL}, 
	from the difference between two results obtained with different $K_{\rm max}$ values.
	However, unlike Ref.~\cite{Ji:2013_PRL}, 
	here we encountered very slow convergence\footnote{
		This may be due to the larger radii of the $A=3$ nuclei compared with $^4$He, 
		and was indeed slightly worse for $^3$He than for $^3$H.
		}, 
	especially with the $\chi$EFT potential, 
	and particularly for the $\delta^{(2)}$ terms, 
	which turned out to be more sensitive to the parameterization of the hyperradial grid. 
	Consequently, 
	the final values we provide for many terms\footnote{
		Depending on the nucleus and the nuclear potential, some of the following terms were obtained with the aid of extrapolations: BE, $\alpha_E$, the individual $\delta^{(2)}$ terms (including $\delta^{(2)}_{NS}$), $\delta^{(0)}_M$, and the sum of all other $\delta^{(0)}$ terms.
		}
	were obtained by extrapolating the results of several calculations made with different $K_{\rm max}$ values. 
	Therefore, for some of our results the numerical uncertainty was estimated from these extrapolations. 
	\item [Nuclear model]
	Next we note the dependence of the results on the nuclear model, 
	which is probed as in Ref.~\cite{Ji:2013_PRL} by using the AV18/UIX and $\chi$EFT potentials. 
	The final values we present are obtained by taking the mean value of the two results. 
	As in Refs.~\cite{Ji:2013_PRL,Hernandez:2014_PLB}, the corresponding uncertainty is estimated as their difference divided by $\sqrt{2}$, to account for the possibility that the ``true'' result lies outside the range bounded by the two calculated results.
	\item [ISB] 
	The next source of uncertainty stems from the conservation of isospin symmetry in our calculations, 
	{\it i.e.}, we assume that the total isospin is a conserved quantity throughout the calculation. 
	All nucleons are taken to be of equal mass, which is the average between the proton and neutron masses. 
	The difference between the proton and the neutron is manifested only in their gyromagnetic factors and in the electromagnetic interaction included in the NN interaction. 
	In the $A=3$ nuclei, most of the isospin symmetry breaking (ISB) effect can be accounted for by allowing the nuclear ground-state wave functions to include also the channel with higher total isospin value $T=3/2$. 
	This, however, increases the number of basis states in each calculation and the associated computational cost rises rapidly with $K_{\rm max}$.  
	It was therefore performed selectively only to estimate the uncertainty associated with performing isospin conserving calculations. 
	\item [Nucleon-size corrections] 
	Comparing the coordinate-space and momentum-space treatments of the NS corrections we conclude that higher-order corrections to the terms we obtained are expected only at $\delta^{(1)}_{NS}$. 
	For the Zemach moments we were able to undertake a more accurate approach, 
	from which we estimate these higher-order corrections to be 
	$\sim\! 1.46$\% ($\sim\! 1.33$\%) for $\mu\,^3$He$^+$ ($\mu\,^3$H). 
	However, for consistency we use here only the low-$Q^2$ approximation of the nucleon electric form factor,
	and use the above corrections to estimate the NS-related uncertainties of the $\delta^{(1)}$ terms.
	\item [Relativistic corrections] 
	As explained in Section~2, 
	relativistic corrections were included only for the leading electric dipole contribution $\delta^{(0)}_{D1}$. 
	Their sum turned out to be $2.0$\% ($2.1$\%) of the non-relativistic value in $^3$He ($^3$H). 
	We therefore estimated the uncertainty due to uncalculated relativistic corrections of the other contributions to be of that relative size.
	We would like to point out two aspects of the elastic (Zemach) term: 
	(i) 
	Comparing the non-relativistic calculation of $\delta^A_{\rm Zem}$ of $\mu$D in Ref.~\cite{Hernandez:2014_PLB} with the relativistic treatment in Ref.~\cite{Carlson:2014_PRA} reveals a discrepancy of the same order as the uncertainty estimate given above. 
	(ii) 
	We calculate $\delta^A_{\rm Zem}$ according to the definition that connects it to the nuclear Zemach moments. 
	Therefore, no relativistic corrections are needed in the comparison we make with similar results in the literature. 
	However, when our value for $\mathcal{C}$ is used to approximate the full elastic part of $\delta^A_{\rm TPE}$ in Eq.~(1), the missing relativistic corrections should be accounted for. In this context, the total relative uncertainty of $\mathcal{C}$ should be increased to $2.5$\%. 
	\item [Coulomb corrections]
	Following Ref.~\cite{Pachucki:2015_PRA}, we estimate higher-order Coulomb corrections to be $\sim$6\% of $\delta^{(2)}$.
	\item [Multipole expansion] % eta expansion 
	As in Ref.~\cite{Ji:2013_PRL}, our multipole expansion is truncated at $\delta^{(2)}$. 
	Based on our results we conservatively estimate 2\% uncertainty in $\delta^A_{\rm pol}$ due to this truncation. 
	\item [$\bs {Z\alpha}$ expansion] 
	Except for the logarithmically enhanced Coulomb distortion contribution, 
	we include in our calculation of $\delta^A_{\rm pol}$ all terms of order $(Z\alpha)^5$. 
	Since $(Z\alpha)$ is small for these systems, the missing contribution from all the higher-order terms can be approximated by the first unaccounted-for term in the series, $(Z\alpha)^6$, 
	which is naturally estimated to be $(Z\alpha) \simeq$1.46\% (0.73\%) of the size of $\delta^A_{\rm pol}$ in $^3$He ($^3$H). 
	\item [Magnetic MEC contribution]
	The magnetic dipole term $\delta^{(0)}_{M}$ is calculated using the impulse approximation (IA) operator, 
	and is therefore missing significant corrections, mainly due to meson-exchange currents (MECs)\footnote{
		See Refs.~\cite{Marcucci:2008_PRC,Pastore:2013_PRC} and Refs.~therein.
		}.
	The same IA operator was also used to calculate the magnetic moment $\hat{\mu}_{gs}$ in the nuclear ground state. 
	The results, presented in Table~I, 
	show a deviation of the IA $\hat{\mu}_{gs}$ calculated with the AV18/UIX ($\chi$EFT) potential from the very precise experimental values by 23\% (21\%) for $^3$He and 15\% (13\%) for $^3$H. 
	The same relative errors were therefore assumed also for the small IA value obtained for $\delta^{(0)}_{M}$.
\end{description}

The total uncertainties were obtained as a quadrature sum of all the above, 
where the last five sources do not affect the Zemach terms. 
The values we thus obtained for the individual and total relative uncertainties estimated for $\delta^A_{\rm pol}$, $\delta_{\rm Zem}^{A}$, and $\delta^A_{\rm TPE}$ in $\mu\,^3$He$^+$ and $\mu\,^3$H are given in Table~\ref{table_err_supp} below. 
We remind the reader that $\delta^A_{\rm TPE} \equiv \delta_{\rm Zem}^{A}+\delta^A_{\rm pol}$ can be obtained directly from our results, by summing all terms in $\delta^A_{\rm pol}$ except for the Zemach terms, due to their cancellation.

% Uncertainty Estimates
%=======================
\begin{table}[htb]
	\caption{Estimated relative uncertainties, in percents, 
		assigned to the calculated nuclear TPE corrections 
		to the \mbox{2S-2P} Lamb shift in $\mu\,^3$He$^+$ and $\mu\,^3$H. 
		The presented values are rounded. 
		The total uncertainties are obtained from a quadrature sum.
		}
	\label{table_err_supp}
	\begin{center}
		\footnotesize
		%\scriptscriptstyle
		\renewcommand{\tabcolsep}{1.2mm}
		\begin{tabular}{@{} l  l l l    c    l l l @{}}
			\hline\hline
			& \multicolumn{3}{c}{$\ \bs{\mu\,^3{\mathrm{He}}^+}$} 
			& & \multicolumn{3}{c}{$\bs{\mu\,^3}$\textbf{H}} \\ 
			Error type &
			$\delta^A_{\rm pol}\ $& $\ \ \delta_{\rm Zem}^{A}$& $\delta^A_{\rm TPE}$&& 
			$\delta^A_{\rm pol}\ $& $\ \ \delta_{\rm Zem}^{A}$& $\delta^A_{\rm TPE}$\\
			\hline
			%                                  3He           |                  3H
			% error type            pol     Zem        TPE   |      pol      Zem       TPE
			%
			Numerical             & 0.4   & 0.1      & 0.1   &$\ $& 0.1    & 0.0      & 0.1 \rule{0pt}{3.5ex}\\
			Nuclear model		  & 1.5   & 1.8      & 1.7   &$\ $& 2.2    & 2.3      & 2.2 \rule{0pt}{3.5ex}\\
			ISB                   & 2.0   & 0.2      & 0.5   &$\ $& 0.9    & 0.2      & 0.6 \rule{0pt}{3.5ex}\\
			Nucleon size          & 1.6   & 1.5      & 0.6   &$\ $& 0.6    & 1.3      & 0.0 \rule{0pt}{3.5ex}\\
			Relativistic          & 0.6   & $\quad$- & 1.5   &$\ $& 1.4    & $\quad$- & 0.3 \rule{0pt}{3.5ex}\\
			Coulomb		          & 1.2   & $\quad$- & 0.3   &$\ $& 0.3    & $\quad$- & 0.2 \rule{0pt}{3.5ex}\\
			Multipole expansion   & 2.0   & $\quad$- & 0.6   &$\ $& 2.0    & $\quad$- & 1.4 \rule{0pt}{3.5ex}\\
			Higher $Z \alpha$     & 1.5   & $\quad$- & 0.4   &$\ $& 0.7    & $\quad$- & 0.5 \rule{0pt}{3.5ex}\\
			Magnetic MEC          & 0.4   & $\quad$- & 0.1   &$\ $& 0.3    & $\quad$- & 0.2 \rule{0pt}{3.5ex}\\
			\hline 
			Total                 & 4.1\% & 2.3\%    & 2.5\% &$\ $& 3.6\%  & 2.7\%    & 2.7\% \rule{0pt}{3.5ex}\\
			\hline\hline
		\end{tabular}
	\end{center}
\end{table}

\newpage
%%%%%%%%%%%%%%%%%%%%%%%%%%%%%%%%%
% delta^A_pol Table
%%%%%%%%%%%%%%%%%%%%%%%%%%%%%%%%%
\section{List of individual contributions to the nuclear polarization energy correction}
\begin{table}[htb]
	\caption{Nuclear structure corrections to the \mbox{2S-2P} Lamb shift $\Delta E$ [meV] in $\mu\,^3$He$^+$ and $\mu\,^3$H,
		obtained with the AV18/UIX and $\chi$EFT nuclear potentials. 
		The brackets show only the numerical error in the presented precision. 
		See text for details regarding the individual terms.}
	\label{table_pol}
	\begin{center}
		\footnotesize
		\renewcommand{\tabcolsep}{1.0mm}
		\begin{tabular}{@{} c @{} l   r   c   r  c  r @{} c  r @{}}
			\hline\hline
			& & \multicolumn{3}{c}{$\ \bs{\mu\,^3{\mathrm{He}}^+}$} 
			& & \multicolumn{3}{c}{$\bs{\mu\,^3}$\textbf{H}} \\ 
			& & \multicolumn{2}{r}{AV18/UIX}$\ \ $& $\chi$EFT$\quad $ 
			& & \multicolumn{2}{r}{AV18/UIX}$\quad $& $\chi$EFT$\quad $ \\ 
			\hline
			%                                                       3He                         3H
			%                                           AV18/UIX         EFT:EM+N        AV18/UIX    EFT:EM+N
			$\delta^{(0)}$   & $\delta^{(0)}_{D1}$   &${-6.479(06)}$& &${-6.633(1)}$&&${-0.7669(6)}$& &${-0.7848(1)}$\\
			\rule{0pt}{3.5ex}& $\delta^{(0)}_{L}$    &  0.232(00)   & &  0.240(0)   &&  0.0285(0)   & &  0.0296(0)\\
			\rule{0pt}{3.5ex}& $\delta^{(0)}_{T}$    &${-0.103(00)}$& &${-0.107(0}$ &&${-0.0128(0)}$& &${-0.0132(0)}$\\
			\rule{0pt}{3.5ex}& $\delta^{(0)}_{C}$    &  1.000(01)   & &  1.020(3)   &&  0.0718(1)   & &  0.0732(0)\\
			%
			%\rule{0pt}{3.5ex}& 
			sum &             					     &${-5.346(05)}$& &${-5.486(7)}$&&${-0.6788(5)}$& &${-0.6956(3)}$\\
			\\ %\hline
			%\rule{0pt}{3.5ex}
			& $\delta^{(0)}_{M}$    &  0.081(02)   & &  0.047(0)   &&  0.0101(3)   & &  0.0058(0)\\
			\\ %\hline
			$\delta^{(1)}$   & $\delta^{(1)}_{R3}$ &${-8.539(12)}$& &${-8.711(3)}$&&  -$\qquad$   & &  -$\qquad$\\
			\rule{0pt}{3.5ex}& $\delta^{(1)}_{Z3}$ &  8.100(10)   & &  8.327(3)   &&  0.1778(0)   & &  0.1844(0)\\
			\\ %\hline
			$\delta^{(2)}$   & $\delta^{(2)}_{R^2}$  &  0.632(00)   & &  0.654(3)   &&  0.0199(0)   & &  0.0206(2)\\
			\rule{0pt}{3.5ex}& $\delta^{(2)}_{Q}$    &  1.015(01)   & &  1.038(0)   &&  0.0344(0)   & &  0.0358(0)\\
			\rule{0pt}{3.5ex}& $\delta^{(2)}_{D1D3}$ &${-0.841(00)}$& &${-0.862(0)}$&&${-0.0783(1)}$& &${-0.0811(1)}$\\
			\\ %\hline
			$\delta_{NS}$    & $\delta^{(1)}_{R1}$   &${-1.294(00)}$& &${-1.314(0)}$&&  0.0280(0)   & &  0.0287(0)\\
			\rule{0pt}{3.5ex}& $\delta^{(1)}_{Z1}$   &  2.256(01)   & &  2.291(0)   &&  0.0453(0)   & &  0.0463(0)\\
			\rule{0pt}{3.5ex}& $\delta^{(2)}_{NS}$   &${-0.179(00)}$& &${-0.185(0)}$&&${-0.0272(0)}$& &${-0.0282(0)}$\\
			\hline %\hline
			\multicolumn{2}{l}{$\delta^A_{\rm pol}$} & 
			\rule{0pt}{3.5ex}                         ${-4.114(17)}$& &${-4.201(8)}$&&${-0.4688(6)}$& &${-0.4834(4)}$\\
			\multicolumn{2}{l}{$\delta_{\rm Zem}^{A}$}   & 
			\rule{0pt}{3.5ex}                        ${-10.356(10)}$& &${-10.618(3)}$&&${-0.2232(0)}$& &${-0.2307(0)}$\\
			{$\delta^A_{\rm TPE}$} && 
			\rule{0pt}{3.5ex}                      ${-14.470(14)}$  & &${-14.819(8)}$&&${-0.6920(6)}$& &${-0.7140(4)}$\\
			\hline\hline
			\end{tabular}
	\end{center}
\end{table}

%%%%%%%%%%%
% The END %
%%%%%%%%%%%
\end{document}